\documentclass[runningheads]{llncs}

\usepackage[FINAL,year=2024,ID=6635]{eccv}
\usepackage{eccvabbrv}

\usepackage{graphicx}
\usepackage{booktabs}
\usepackage{amsmath}
\usepackage{amsfonts}
\usepackage{amssymb}
\usepackage{bm}
\usepackage{multirow}
\usepackage{booktabs}

\newcommand{\ToolName}{WeConvene}

\usepackage[accsupp]{axessibility}  % Improves PDF readability for those with disabilities.

\usepackage[pagebackref,breaklinks,colorlinks,citecolor=eccvblue]{hyperref}
\usepackage{orcidlink}

\begin{document}

\title{\ToolName{}: Learned Image Compression with \underline{W}av\underline{e}let-Domain \underline{Conv}olution and \underline{En}tropy Mod\underline{e}l} 

\titlerunning{WeConvene}

\author{Haisheng Fu\inst{1} \and
Jie Liang\inst{1}\and
Zhenman Fang \inst{1} \and
Jingning Han \inst{2} \and
Feng Liang \inst{3} \and
Guohe Zhang \inst{3}}

\authorrunning{H.~Fu et al.}

\institute{School of Engineering Science, Simon Fraser University, Canada  \and Google LLC, USA
 \and School of Microelectronics, Xi'an Jiaotong University, China}
\maketitle
\begin{abstract}
Recently learned image compression (LIC) has achieved great progress and even outperformed the traditional approach using DCT or discrete wavelet transform (DWT). However, LIC mainly reduces spatial redundancy in the autoencoder networks and entropy coding, but has not fully removed the frequency-domain correlation explicitly as in DCT or DWT. To leverage the best of both worlds, we propose a surprisingly simple but efficient \ToolName{} framework, which introduces the DWT to both the convolution layers and entropy coding of CNN-based LIC. First, in both the core and hyperprior autoencoder networks, we propose a \underline{W}av\underline{e}let-domain \underline{C}onvolution (WeConv) module, which performs convolution after DWT, and then converts the data back to spatial domain via inverse DWT. This module is used at selected layers in a CNN network to reduce the frequency-domain correlation explicitly and make the signal sparser in DWT domain. We also propose a \underline{W}av\underline{e}let-domain \underline{Ch}annel-wise \underline{A}uto-\underline{R}egressive entropy \underline{M}odel (WeChARM), where the output latent representations from the encoder network are first transformed by the DWT, before applying quantization and entropy coding, as in the traditional paradigm. Moreover, the entropy coding is split into two steps. We first code all low-frequency DWT coefficients, and then use them as prior to code high-frequency coefficients. The channel-wise entropy coding is further used in each step. By combining WeConv and WeChARM, the proposed WeConvene scheme achieves superior R-D performance compared to other state-of-the-art LIC methods as well as the latest H.266/VVC. For the Kodak dataset and the baseline network with $-0.4\%$ BD-Rate saving over H.266/VVC, introducing WeConv with the simplest Haar transform improves the saving to $-4.7\%$. This is quite impressive given the simplicity of the Haar transform. Enabling Haar-based WeChARM entropy coding further boosts the saving to $-8.2\%$. When the Haar transform is replaced by the 5/3 or 9/7 wavelet, the overall saving becomes $-9.4\%$ and $-9.8\%$ respectively. The standalone WeConv layer can also be used in many other computer vision tasks beyond image/video compression. The source code is available at \url{https://github.com/fengyurenpingsheng}.

\end{abstract}
\section{Introduction}
\label{sec:intro}

In the last few years, learned image compression (LIC) methods have quickly outperformed the traditional approaches in both subjective and objective metrics. Linear transform such as the discrete cosine transform (DCT) and discrete wavelet transform (DWT) is a key component in the traditional paradigm, followed by quantization and entropy coding in the transform domain. In LIC, the linear transform is replaced by deep learning-based neural networks, which can be more powerful than linear transform in learning the compact latent representations of the images.

Earlier LIC designs were mainly based on convolutional neural networks (CNN) \cite{Variational,Joint,GLLMM, Lee_2020, Lee_2021, He_2021_CVPR, He_2022_CVPR, jiang2023mlic}. Recently the transformer network has been introduced \cite{Liu_2023_CVPR, zhu2022transformerbased, Qian2022_ICLR}, which can achieve better rate-distortion (R-D) performance, but transformer-based schemes are more difficult to train and have higher requirements on the GPU. These neural networks are also used in the entropy coding part to learn the distributions of the latent representations. As a result, many LIC schemes can get better performance than the traditional approaches, including intra coding in the latest H.266/VVC video coding standard.

Despite its great success, a major limitation of the state-of-the-art LIC schemes is that they do not explicitly remove the frequency-domain redundancy of the latent representations. Although there are some efforts in introducing transform-domain processing to the LIC \cite{Akyazi_2019_CVPR_Workshops,gao2021neural,Mahammand_AAAI,Lin_MMSP,Mahammand_media,Fu_octave,Ma_2022_PAMI,Iliopoulou2023,Zafari_2023}, their performances are not satisfactory.

%\cite{Contextformer_2022, GLLMM, jiang2023mlic, Liu_2023_CVPR}, 

In this paper, we propose a surprisingly simple but efficient way of using DWT in both the autoencoder network and entropy coding parts of the LIC framework, and demonstrate that DWT can indeed significantly improve the performance in the learned image compression paradigm, as expected from the experience in the traditional approach.

Our contributions are summarized as follows:

\begin{figure*}[!thp]
	\centering
		\includegraphics[scale=0.40]{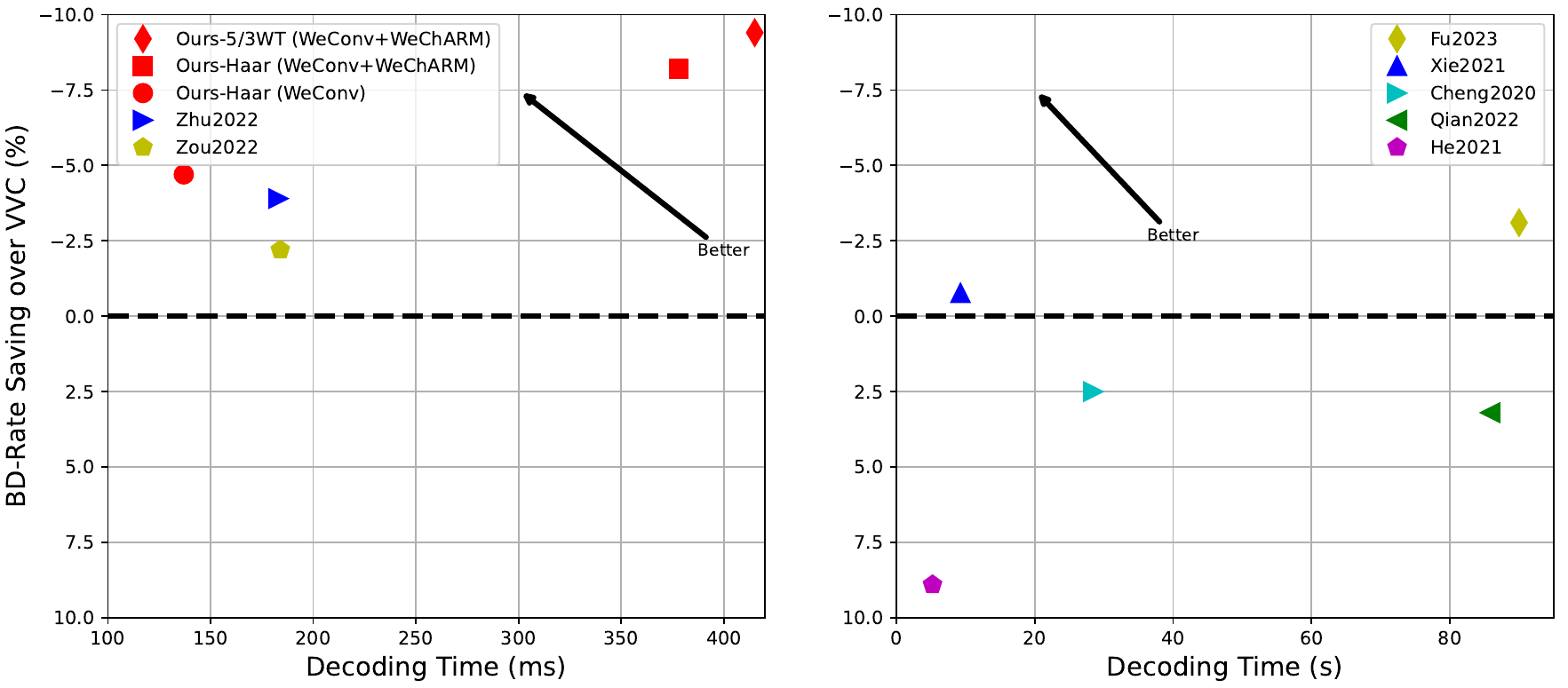}
	\caption{The decoding time and BD-Rate reductions over H.266/VVC for different LIC schemes on the Kodak dataset.}
	\label{BD_rate}
\end{figure*}

\begin{itemize}

\item We propose an effective, low-cost, modular, and plug-and-play WeConv layer, which embeds the convolution between DWT and IDWT, so that the module can still be one layer of a large CNN network. This allows it to enjoy the benefits of CNNs, and also improve the sparsity in the DWT domain. For the Kodak dataset and the baseline network with $-0.4\%$ BD rate saving over H.266/VVC, introducing WeConv with the simplest Haar transform can improve the saving to $-4.7\%$, with negligible change of model size and running time. 

\item  We propose a wavelet domain quantization and entropy coding, denoted as WeChARM, which can explicitly benefit from the improved sparsity given by the WeConv module. For the Kodak dataset, combining Haar-based WeConv and WeChARM entropy coding further boosts the saving to $-8.2\%$, with moderate increase of model size and running time. When the Haar transform is replaced by the 5/3 or 9/7 wavelet, the overall saving can be improved to $-9.4\%$ and $-9.8\%$ respectively. 

\item Since the proposed scheme is based on CNN, it is easier to train and has less requirements on GPU than transformer-based schemes. It also does not use other high complexity operators such as non-local modules. It therefore achieves a good trade-off between complexity and performance, as shown in Fig. \ref{BD_rate}, establishing our approach as the new state of the art in LIC.

\item We show that with judicious design, the traditional
wavelet transform can be used in LIC and achieve the state of the art. The scheme can be further improved. This opens up many future topics, and will bring in more attentions from the signal processing community. The proposed WeConv module can also be used in other applications beyond image compression.

\end{itemize}

\section{Background and Related Work}

\subsection{Traditional Image Compression Methods}

Traditional image and video codings, such as JPEG \cite{JPEG}, JPEG 2000 \cite{JPEG2000}, and H.264/H.265/H.266, extensively utilize linear transforms such as the Discrete Cosine Transform (DCT) and Discrete Wavelet Transform (DWT) to remove the redundancy in the frequency domain. The transformed data is then quantized to remove small coefficients in the frequency domain without introducing too much reconstruction error. The remaining redundancy is removed using entropy coding.

\subsection{Representative LIC Methods}

In the last few years, learned image compression (LIC) \cite{He_2021_CVPR, He_2022_CVPR, Liu_2023_CVPR, Qian2022_ICLR, zhu2022transformerbased, Asymmetric_Fu, Guo_2022, Fu_ICASSP} has witnessed remarkable advancements, and started to outperform traditional methods, including the latest H.266/VVC. In \cite{end_to_end}, the first end-to-end learned image compression framework was proposed, which outperformed JPEG and JPEG2000 by using a novel architecture of convolutions and nonlinear activation functions. In \cite{Variational}, a variational autoencoder structure with a hyperprior is proposed for capturing spatial dependencies, achieving comparable performance to BPG (4:4:4). In \cite{Joint}, based on \cite{Variational} and by combining autoregressive and hierarchical priors, the method was able to beat BPG (4:4:4) on both PSNR and MS-SSIM metrics.

Based on \cite{Joint}, some LIC methods \cite{Lee_2021, GLLMM, Contextformer_2022, xie2021enhanced} utilize serial context-adaptive models to achieve better performance than H.266/VVC. However, serial context models are time-consuming. To address this issue, a channel-wise autoregressive entropy model (ChARM) is introduced in \cite{channel} to avoid element-level serial processing. Furthermore, in \cite{He_2022_CVPR}, a spatial-channel contextual adaptive model is proposed to speed us the entropy coding without compromising the R-D performance. Similarly, in \cite{He_2021_CVPR}, a checkerboard context model (CCM) is developed to improve parallelism, but the R-D performance is reduced slightly compared to serial context model.

In \cite{cheng2020}, efficient residual network is proposed to extract more compact and efficient latent representation. Attention modules are also used, which has been adopted in some other schemes \cite{Zou_2022_CVPR,Contextformer_2022,gao2021neural}. In \cite{xie2021enhanced}, the invertible neural networks (INNs) are used to enhance overall performance.

Recently the transformer network has been introduced to LIC \cite{Liu_2023_CVPR, zhu2022transformerbased, Qian2022_ICLR}. For example, in \cite{Liu_2023_CVPR},  the swin-transformer is combined with a ChARM model to enhance spatial dependency capture. However, transformer-based schemes are more difficult to train and have higher requirements on the GPU.

%\subsection{Attention-based Learned Image Compression Methods}

\subsection{Efforts in Frequency-domain LIC}

There have been some efforts in introducing frequency-domain processing to the LIC.

In \cite{Akyazi_2019_CVPR_Workshops}, the authors applied “3-scale Daubechies-1” wavelets, and then introduced various CNN layers in the wavelet domain. The IDWT is only used at the end of the decoder. However, the performance was 5-6dB lower than JPEG, and 8-9 dB lower than JPEG2000. A similar idea was used in \cite{Iliopoulou2023}, where the 9/7 wavelet and the network in \cite{Variational} are used, but its results are also not satisfactory.

 In \cite{Mahammand_AAAI,Lin_MMSP,Mahammand_media,Fu_octave}, the octave convolution proposed in \cite{octave} is introduced in LIC, where the feature map in each layer is divided into a low-resolution part and a high-resolution part. The multi-resolution concept is similar to wavelet transform, but the learned filters for the two parts do not necessarily have high-pass or low-pass frequency responses.
 
In \cite{gao2021neural}, the image is decomposed into a low-frequency (LF) part and the residual high-frequency (HF) part, via simple pooling and subtraction operators, similar to Laplacian pyramid. The two parts are processed separately and merged by a dual attention module. In \cite{Zafari_2023}, the idea is applied to a transformer-based LIC, where the heads in the multi-head self-attention module in the transformer are split into HF heads and LF heads, using pooling and subtraction.

In \cite{Ma_2022_PAMI}, the lifting structure in the wavelet transform is imposed by the neural network architecture, and the filters in each lifting step are learned via training, but it does not use the existing wavelet coefficients, such as 5/3 and 9/7 wavelets.

% However, all the schemes above only borrow some concepts from wavelet transform, but there is no efficient LIC scheme that can directly apply the traditional wavelet transform and enjoy its performance gain as shown in the traditional image coding.
\section{\ToolName{}: LIC with Wavelet-Domain Convolution and Entropy Model}

In this section, we describe the entire architecture of the proposed method, the proposed new components of WeConv and WeChARM, the loss function, and the training of the system.

\begin{figure*}[!thp]
	\centering
		\includegraphics[scale=0.38]{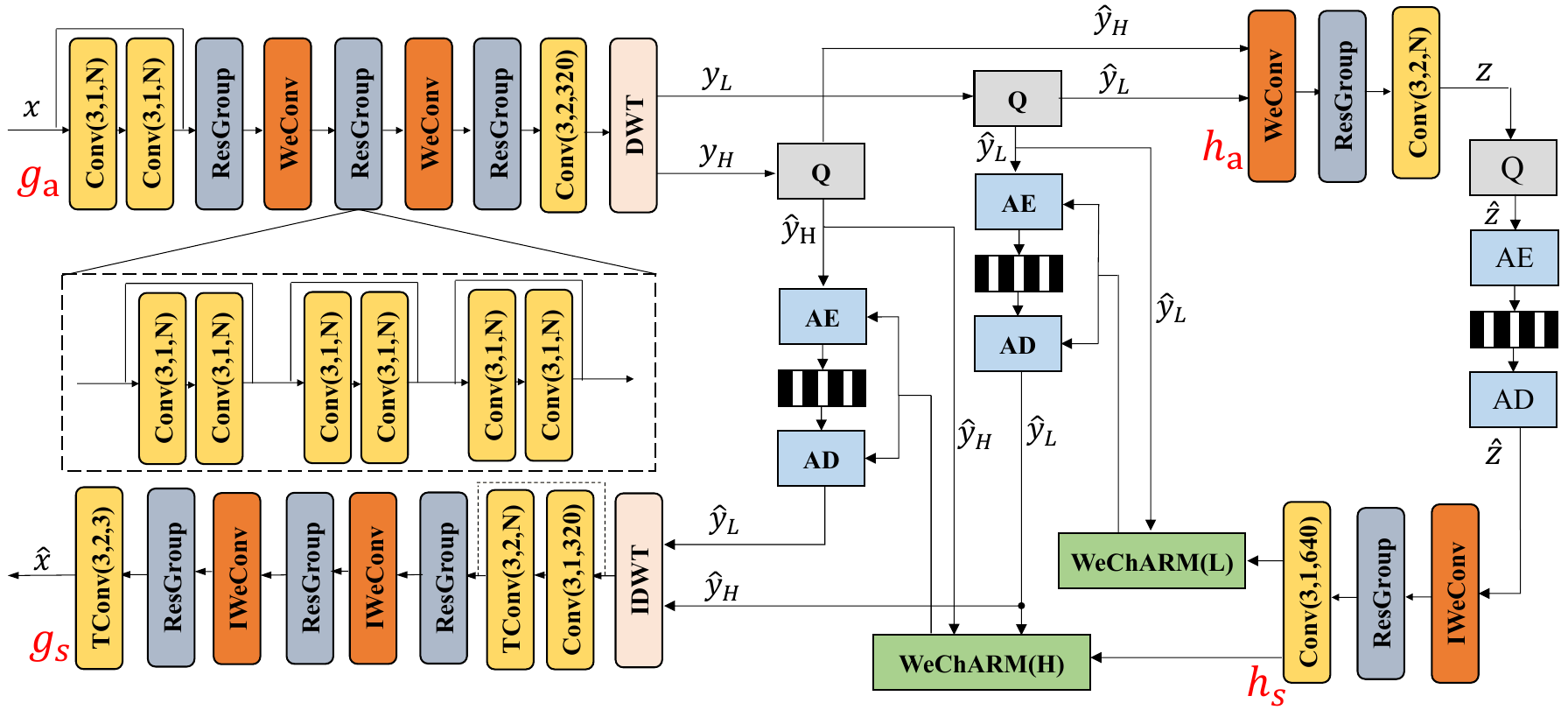}
	\caption{The architecture of the proposed WeConvene scheme. Conv(3, s, n) represents a convolution layer with $3\times3$ kernel size, stride $s$, and $n$ filters. TConv(3, s, n) is the transposed convolution. Dashed shortcut connections represent change of tensor size.  $AE$ and $AD$ stand for Arithmetic Encoder and Arithmetic Decoder, respectively.}
	\label{whole_networkstructure}
\end{figure*}

The proposed scheme is depicted in Fig. \ref{whole_networkstructure}. As in other popular LIC methods, our system includes the core autoencoder $g_a$ to extract the compact latent representations of the input image, the core decoder network $g_s$ to reconstruct the image, the hyperprior networks $h_a$ and $h_s$ to encode and decode the side information that helps the entropy coding of the latents.

The input color image has dimension $W \times H \times 3$. The pixel values are normalized to the range of $[-1, 1]$. The encoder network $g_a$ includes multiple layers of convolutions and leaky ReLU. Some layers are grouped into ResGroup modules, each includes three residual blocks, as shown in Fig. \ref{whole_networkstructure}. Some layers use the proposed WeConv module, which includes the pooling or downsampling operation, and will be described in Sec. \ref{sec_Farm}.

Another contribution of our scheme is to apply the DWT at the end of the core encoder network to convert the latent representations into the wavelet domain. This makes the coefficients sparser and can improve the subsequent quantization and entropy coding.

The wavelet-domain coefficients are then quantized. To reduce the bit rate, the entropy coding is divided into two steps. The LF quantized DWT subband $\hat{y}_L$ is first encoded, which is then used to encode/decode the three quantized HF DWT subbands $\hat{y}_H$. 

As in \cite{channel,Liu_2023_CVPR}, we use the fast channel-wise entropy coding (ChARM) to encode the LF and HF coefficients, denoted by WeChARM (L) and WeChARM (H) in Fig. \ref{whole_networkstructure}. The details of WeChARM will be explained in Sec. \ref{conditioned_conding}.

As in other LIC schemes, to improve the entropy coding performance, the hyperprior networks $h_a$ and $h_s$ are utilized to encode and decode additional prior information $z$ for both $y_L$ and $y_H$. The WeConv module is also used in the hyperprior encoding networks. The inverse WeConv (IWeConv) module with transposed convolution is used in the core decoding network and the hyperprior decoding network, as described in Sec. \ref{sec_Farm}.

%%The final step involves reconstructing the image using the core decoder, leveraging advancements to accelerate the context model processing.  Following prior works \cite{Liu_2023_CVPR, He_2022_CVPR}, we encode each $(\left| y - \mu \right| + \mu)$ to the bitstream instead of $ y $ and restore the coding-symbol as $(\left| y - \mu \right| + \mu)$, which can further benefit the single Gaussian entropy model.

%%The design ensures symmetry between the encoders and decoders in both core and hyper networks, differing only in their use of convolution and deconvolution operations, streamlining the overall compression and reconstruction process.

\subsection{\underline{W}av\underline{e}let-domain \underline{C}onvolution (WeConv) Module}
\label{sec_Farm}

\begin{figure}[tb]
  %\centering
  \begin{subfigure}{0.5\linewidth}
    \includegraphics[scale=0.27]{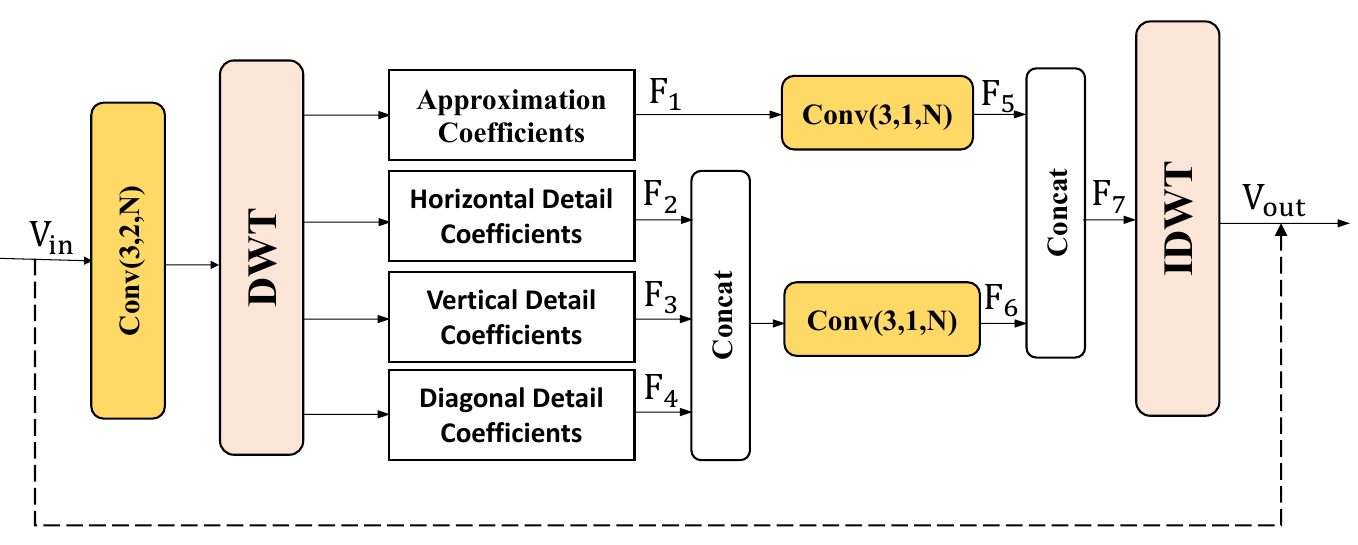}
    \caption{}
    \label{fig:short-a}
  \end{subfigure}
  \hfill
  \begin{subfigure}{0.5\linewidth}
    \includegraphics[scale=0.27]{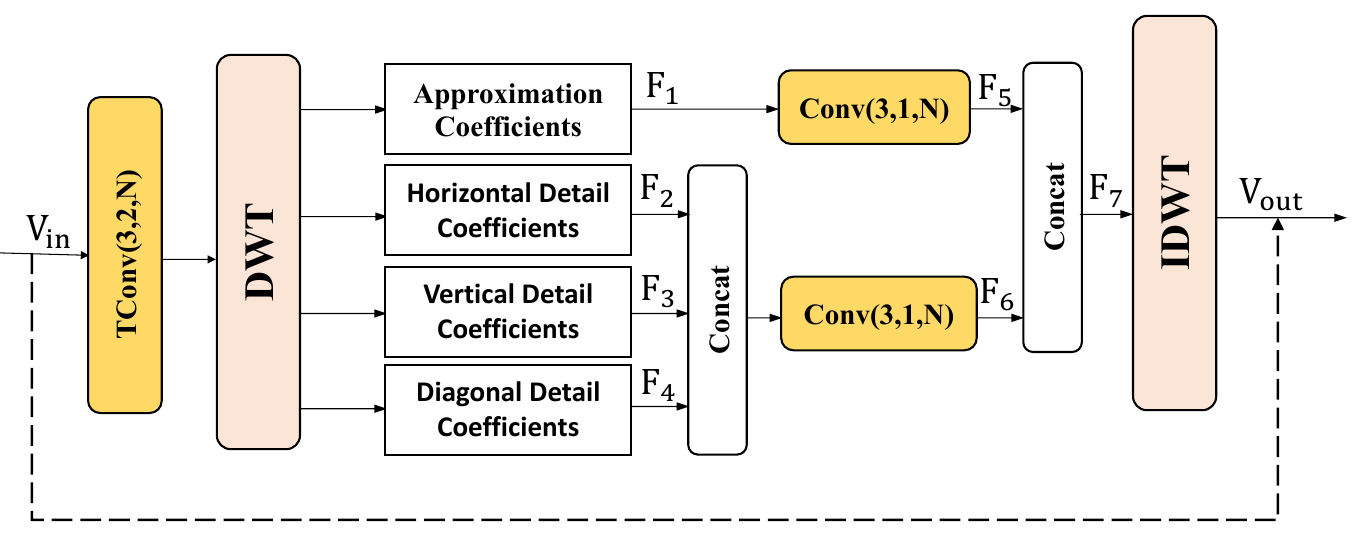}
    \caption{}
    \label{fig:short-b}
  \end{subfigure}
  \caption{(a) The architecture of forward WeConv network with downsampling; (b) The architecture of inverse WeConv (IWeConv) network with upsampling. }
  \label{fig:WeConv}
\end{figure}

Fig. \ref{fig:WeConv} shows the details of the proposed WeConv and inverse WeConv modules. In this paper, we use WeConv when the size of the latent representation is changed (this might not be necessary in other applications). The input signal first passes through a convolutional layer, which also performs downsampling or upsampling, and then converted to the wavelet domain by the DWT operator. 

In this paper, we use the $2 \times 2$ Haar transform, the 5/3 and 9/7 wavelets in JPEG 2000 as examples of the DWT. In Sec. \ref{sec_results}, we will compare the performance of the three wavelets.

After the DWT, the coefficients in the LF subband, $F_1$, are filtered by one set of convolutions. The three HF subbands, $F_2$, $F_3$, $F_4$, are concatenated and filtered by another set of convolutions. 

We then split the HF subbands to their original locations, and apply the inverse DWT to obtain the corresponding spatial-domain latent representations, which can be processed by the subsequent convolution layers as usual. The shortcut connection with different sizes as proposed in the ResNet is applied in the spatial domain.

The structure of the inverse WeConv module is similar to the forward WeConv, except that the transposed convolution is used to upsample the signal. 

The Generalized Divisive Normalization (GDN) is used in WeConv and IWeConv modules, which has better performance than the Leakly ReLU \cite{Variational}.

The proposed WeConv module uses DWT to transform the input data into the wavelet domain, performs subband-based convolution, and then transforms the signal back to the time domain by IDWT. Therefore, it can be used as a standalone layer in CNN networks without drastically disrupting the typical signal distributions in CNNs, which could make it difficult to design a good network, as shown by the unsatisfactory performance in \cite{Akyazi_2019_CVPR_Workshops,Iliopoulou2023}, where the entire CNNs are performed in the wavelet domain.

\subsection{\underline{W}av\underline{e}let-domain \underline{Ch}annel-Wise \underline{A}uto-\underline{R}egressive Entropy \underline{M}odel (WeChARM)}

\label{conditioned_conding}

\begin{figure*}[!thp]
	\centering
		\includegraphics[scale=0.37]{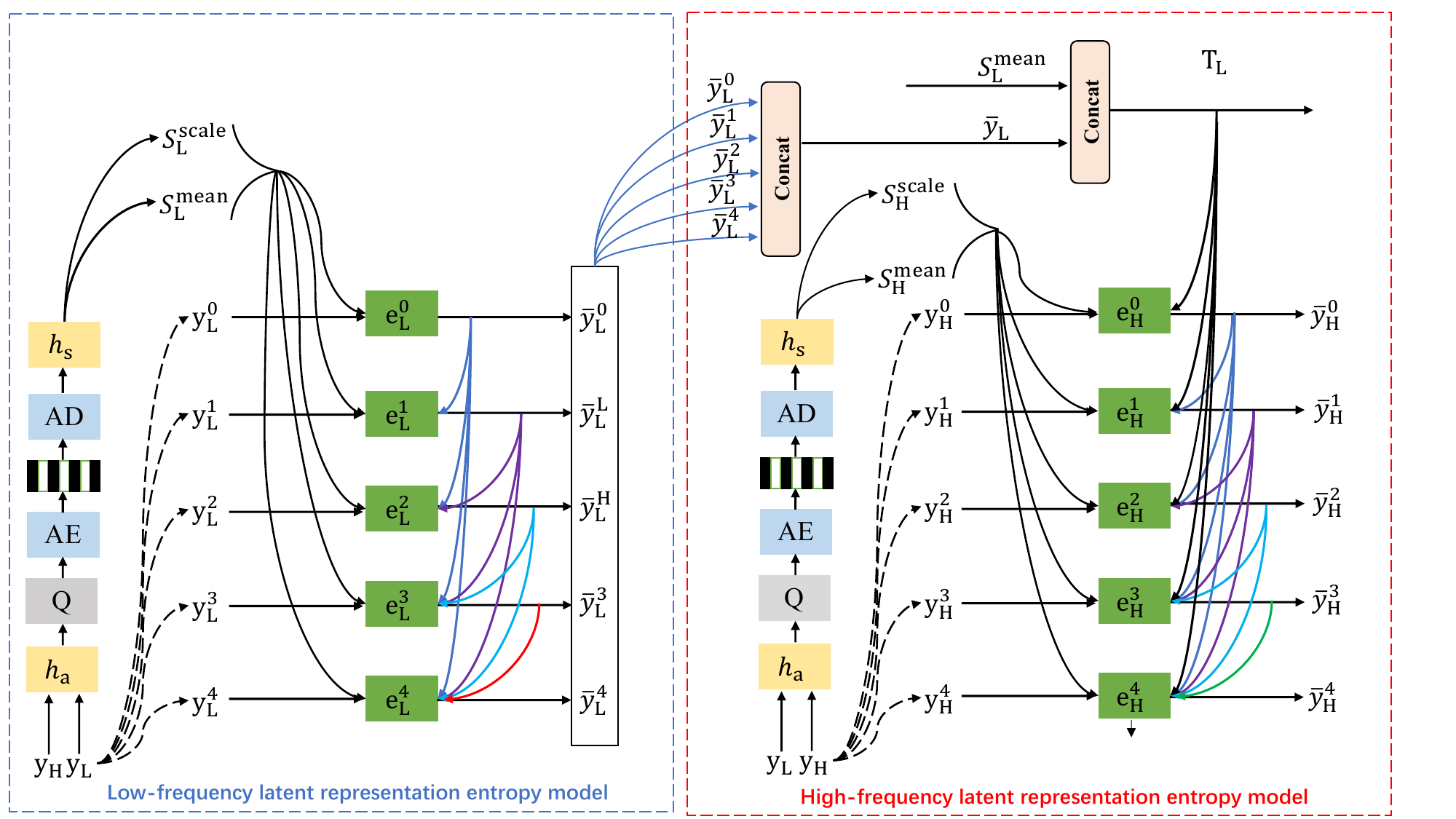}
	\caption{The details of the proposed WeChARM modules for LF and HF subbands.}
	\label{channel_wise_entropy_model}
\end{figure*}

\begin{figure*}[!thp]
	\centering
\includegraphics[scale=0.33]{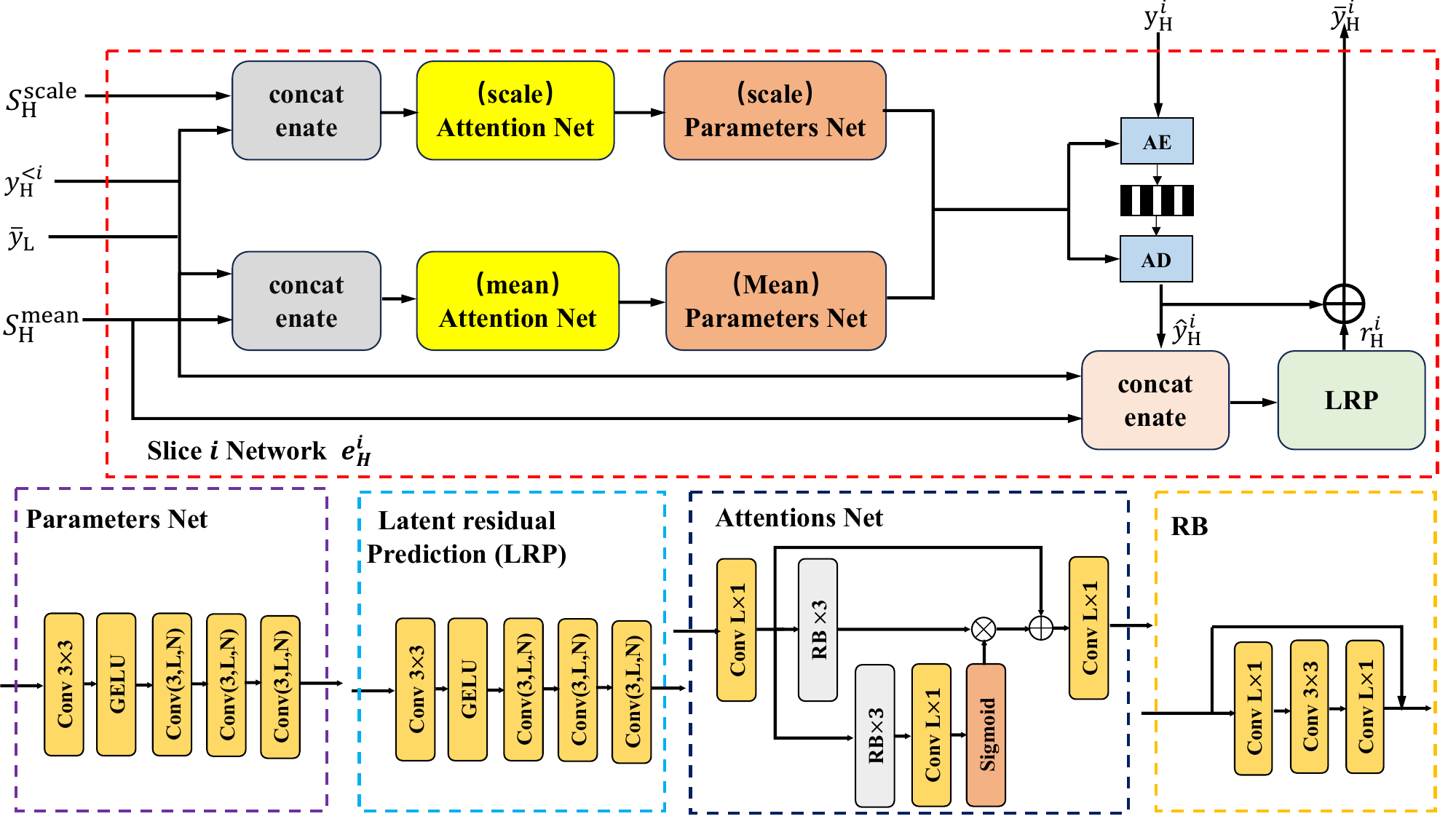}
	\caption{The slice coding network $e_H^i$ for the HF entropy coding in Fig. \ref{channel_wise_entropy_model}.}
	\label{High_frequency_channel_wise_entropy_model_network}
\end{figure*}

In this part, we explain the details of the two WeChARM modules in Fig. \ref{whole_networkstructure} to encode the LF and HF components $y_{L}$ and $y_{H}$ in the wavelet domain, as shown in Fig. \ref{channel_wise_entropy_model} and Fig. \ref{High_frequency_channel_wise_entropy_model_network}.

The channel-wise auto-regressive entropy (ChARM) module was first introduced in \cite{channel}. In \cite{Liu_2023_CVPR}, a Swin-transformer-based attention mechanism (SWAtten) is used. It also reduces the number of slices in \cite{channel} from 10 to 5 to improve the trade-off between running speed and R-D performance. 

In this paper, we apply the ChARM model in \cite{Liu_2023_CVPR} with 5 slices to encode both the LF and HF components $y_{L}$ and $y_{H}$ in the wavelet domain, as shown in Fig. \ref{channel_wise_entropy_model}. Each slice includes 64 channels. Since our probability modeling is learned in the wavelet domain, it is sparser and more efficient than in the spatial domain. As a result, we found that we can remove the time-consuming SWAtten module in \cite{Liu_2023_CVPR} without affecting the R-D performance.

The five LF slices $y_L^i$ are encoded sequentially by five slice coding networks $e_L^i$ ($i=0, ..., 4$), with the help of the side information $S_L^{scale}$ and $S_L^{mean}$ from hyperprior network ($y_L^i$ is assumed to follow a Gaussian distribution), as well as outputs from the preceding slices to reduce inter-slice redundancy.

After the LF components are coded, they are used to code the five HF slices $y_H^i$ via five networks $e_H^i$. 

Fig. \ref{High_frequency_channel_wise_entropy_model_network} illustrates the details of the slice coding network $e_H^i$. The network structure of $e_L^i$ is similar to $e_H^i$, except that there is no prior information from ${y}_{L}$.

\subsection{Loss Function}

Our loss function is to optimize the R-D performance of the system. Let $R$ represent the expected bitstream length, and $D$ denote the reconstruction error between the source and reconstructed images. The trade-off between rate and distortion is regulated by a Lagrange multiplier, $\lambda$. Consequently, the objective cost function is defined as follows:
\begin{equation}\label{total_loss}
\begin{aligned}
 L &=  \lambda D(\bm{x},\hat{\bm{x}})+H(\hat{\bm{y}}_{L})+H(\hat{\bm{y}}_{H})+H(\hat{\bm{z}}), \\
      H(\hat{\bm{z}}) &=  E [-\log_{2}(P_{\hat{\bm{z}}}(\hat{\bm{z}}))],\\
      H(\hat{\bm{y}}_L) &=  E [-\log_{2}(P_{\hat{\bm{y}}_L | \hat{\bm{z}}}(\hat{\bm{y}}_L | \hat{\bm{z}}))],\\
      H(\hat{\bm{y}}_H) &=  E [-\log_{2}(P_{\hat{\bm{y}}_H | \hat{\bm{y}}_L, \hat{\bm{z}}}(\hat{\bm{y}}_H | \hat{\bm{y}}_L, \hat{\bm{z}}))],
\end{aligned}
\end{equation}
where $D(\bm{x},\hat{\bm{x}})$ is the distortion between the original image $\bm{x}$ and reconstructed image $\hat{\bm{x}}$. We utilize the mean squared error (MSE) and multi-scale structural similarity (MS-SSIM) respectively as our optimized metrics to train our networks. $H(\hat{\bm{y}}_L)$, $H(\hat{\bm{y}}_H)$ and $H(\hat{\bm{z}})$ represent the entropies of the LF, HF components and the hyperprior latent representations, respectively. 

\subsection{Model Training}

The training images are obtained from the CLIC \cite{CLIC_test_2021}, LIU4K \cite{LIU_dataset} and Coco datasets \cite{coco_dataset}, and are resized to $2000 \times 2000$ pixels as part of the data augmentation process, which also includes rotation and scaling. We then obtain 160,000 training image patches with a resolution of $480 \times 480$ pixels.

Our models are optimized using MSE and MS-SSIM metrics respectively. For MSE optimization, the $\lambda$ values are selected from the set of 0.0025, 0.0035, 0.0067, 0.013, 0.025, 0.05, each corresponding to a fixed bit rate, with the number of filters ($N$) for the latent features is set as $128$ for all rates. For MS-SSIM metric,  $\lambda$ is set at 5, 8, 16, 32, and 64, with the filter number $N$ remaining at 128. Each model is trained by $1.5 \times 10^{6}$ iterations using the Adam optimizer, with a batch size of 8 and an initial learning rate of $1 \times 10^{-4}$, which is trained every 100,000 iterations after the initial 750,000 iterations.
\section{Experimental Results}
\label{sec_results}

This section evaluates the proposed method against some state-of-the-art LIC methods and traditional image codes, using both the Peak Signal-to-Noise Ratio (PSNR) and MS-SSIM metrics. The LIC methods include Fu2023 \cite{GLLMM}, FuOctave2023 \cite{Fu_octave},  Zhu2022 \cite{zhu2022transformerbased}, Yi2022 \cite{Qian2022_ICLR}, He2022 \cite{He_2022_CVPR}, He2021 \cite{He_2021_CVPR}, Xie2021 \cite{xie2021enhanced}, AkbariAAAI2021 \cite{Mahammand_AAAI},  AkbariTMM202 \cite{Mahammand_media}, Cheng2020 \cite{cheng2020}, Minnen2020 \cite{channel}, and Minnen2018 \cite{Joint}. The traditional methods are H.266/VVC Intra (4:4:4), and H.265/BPG Intra (4:4:4). 

Three popular test sets are selected, namely the Kodak PhotoCD test set \cite{Kodak}  (24 images with $768 \times 512$ or $512 \times 768$ resolution), the Tecnick 100 test set \cite{Tecnick} (100 images with $1200 \times 1200$  resolution), and the CLIC 2021 test set \cite{CLIC_test_2021} (60 images with resolutions ranging from $751 \times 500$ to $2048 \times 2048$). 

To ensure fair comparisons, we retain the Cheng2020 \cite{cheng2020} method by increasing its number of filters $N$ from 192 to 256 for scenarios requiring higher rates, thereby achieving better performance compared to the original results in \cite{cheng2020} results. The results of other methods come from open-source codes or their original papers.

In the propose WeConvene method, we test three different wavelets: the $2 \times 2$ Haar transform, as well as the 5/3 wavelet and the 9/7 wavelet used in JPEG 2000. Symmetric extension is used at the boundary to avoid boundary artifact and improve the sparsity.

\subsection{R-D Performance}

\begin{figure*}[tb]
  %\centering
  \begin{subfigure}{0.50\linewidth}
    \includegraphics[scale=0.35]{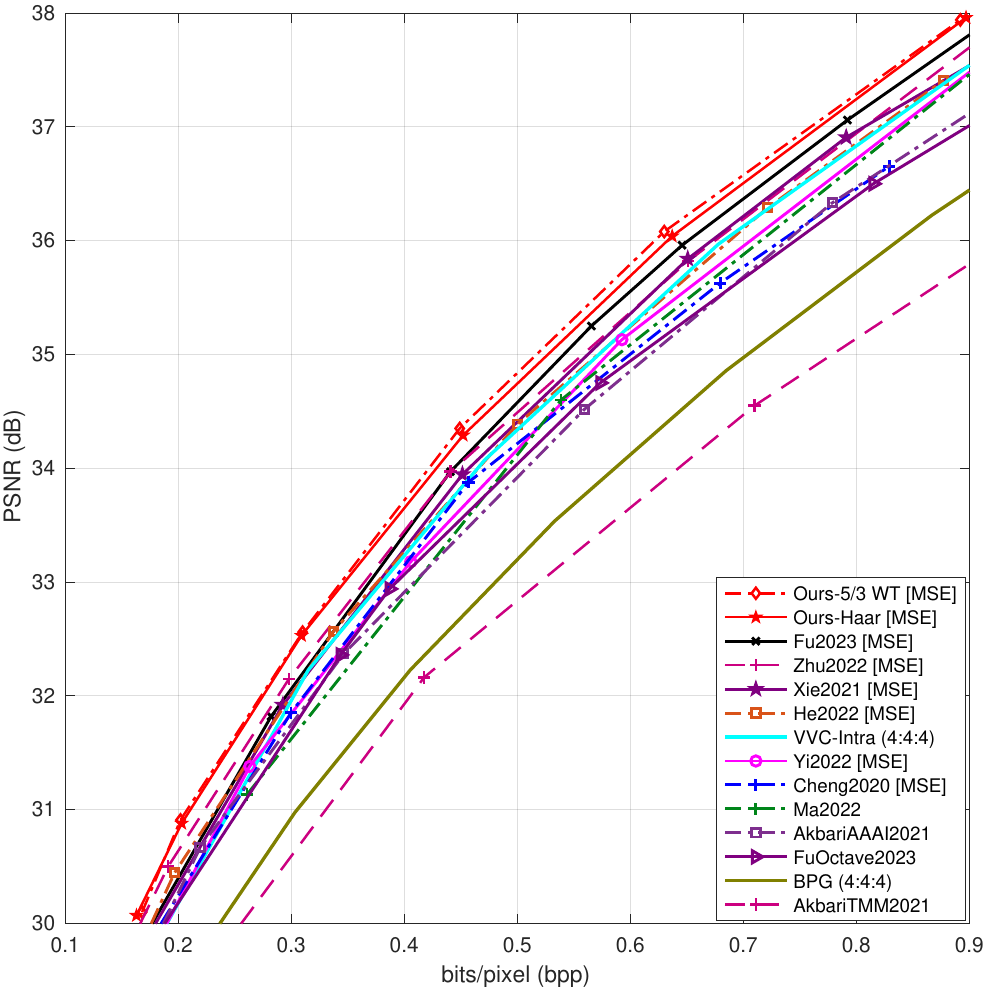}
    \caption{}
    \label{fig:short-a}
  \end{subfigure}
  \hfill
  \begin{subfigure}{0.50\linewidth}
    \includegraphics[scale=0.35]{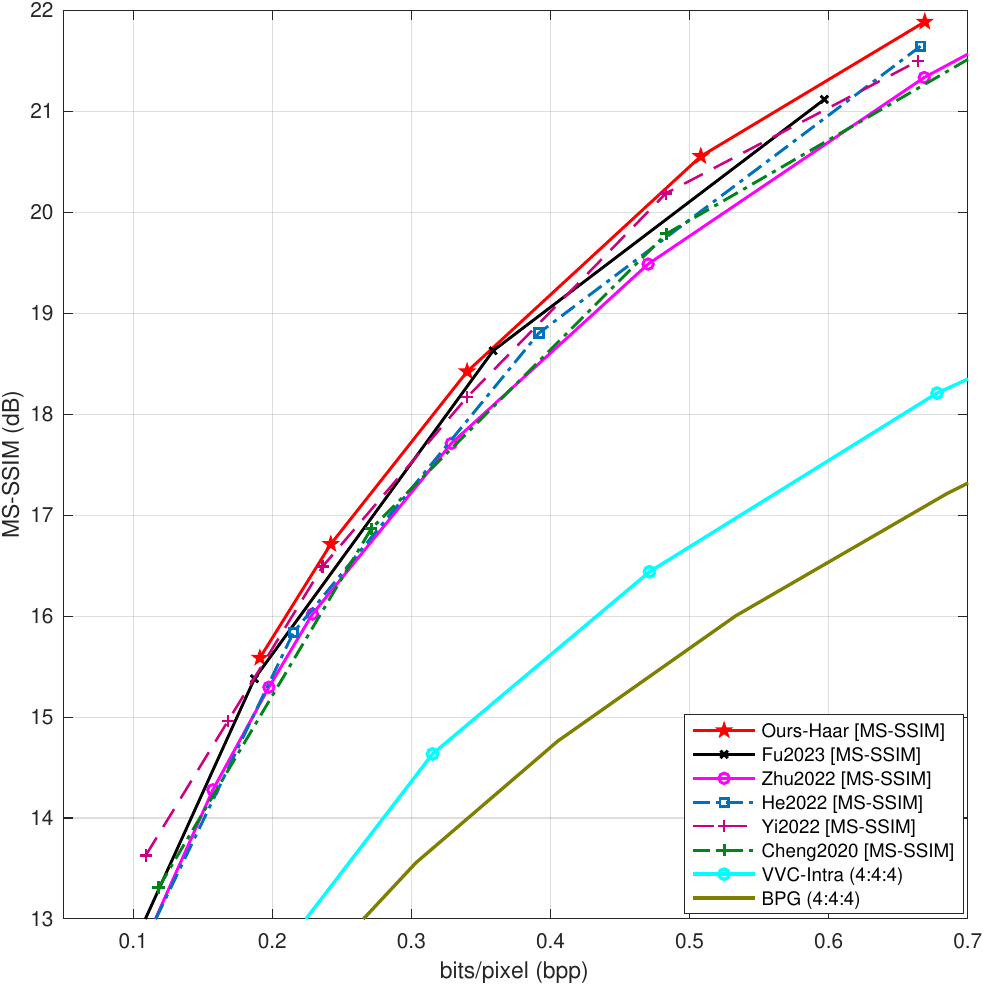}
    \caption{}
    \label{fig:short-b}
  \end{subfigure}
  \caption{The average PSNR (a) and MS-SSIM (b) performances of different methods in the Kodak test set.}
  \label{fig:kodak}
\end{figure*}

Fig. \ref{fig:kodak} depicts the average R-D curves of different methods in all images of the Kodak dataset in terms of PSNR and MS-SSIM metrics. Among the other PSNR-optimized methods, Zhu2022 (MSE) \cite{zhu2022transformerbased} achieves the best performance when the bit rate is lower than 0.43 bpp. It is also better than H.266/VVC at all rates. When the bit rate is higher than 0.43 bpp, Fu2023 (MSE) \cite{GLLMM} achieves the best performance. 

Our proposed WeConvene method with the simple Haar transform consistently outperforms the best of Zhu2022 \cite{zhu2022transformerbased} and Fu2023 \cite{GLLMM} by about $0.2$ dB at all rates, and has a gain of more than $0.5$ dB over VVC, especially at low rates. This is quite impressive given the simplicity of the Haar transform.

When the 5/3 wavelet is used, our performance can be further improved by up to 0.10 dB. The performance of 9/7 wavelet is very similar to 5/3 wavelet, as shown in Fig. \ref{abalation} (b) later. Therefore the result of 9/7 wavelet is not shown in Fig. \ref{fig:kodak}.

In the MS-SSIM metric in Fig. \ref{fig:kodak} (b), our method with the Haar transform also achieves the best performance. Better results can be expected using 5/3 and 9/7 wavelets.

\begin{figure*}[tb]
  %\centering
  \begin{subfigure}{0.50\linewidth}
    \includegraphics[scale=0.35]{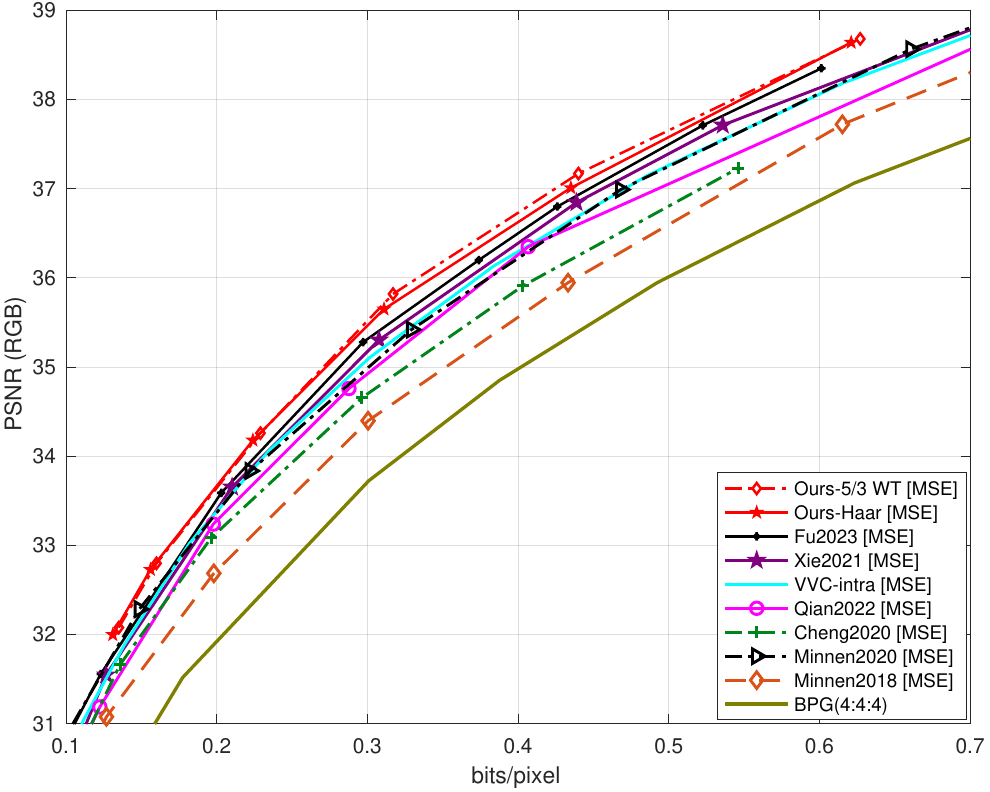}
    \caption{}
    %\label{fig:tecnick-a}
  \end{subfigure}
  \hfill
  \begin{subfigure}{0.50\linewidth}
    \includegraphics[scale=0.35]{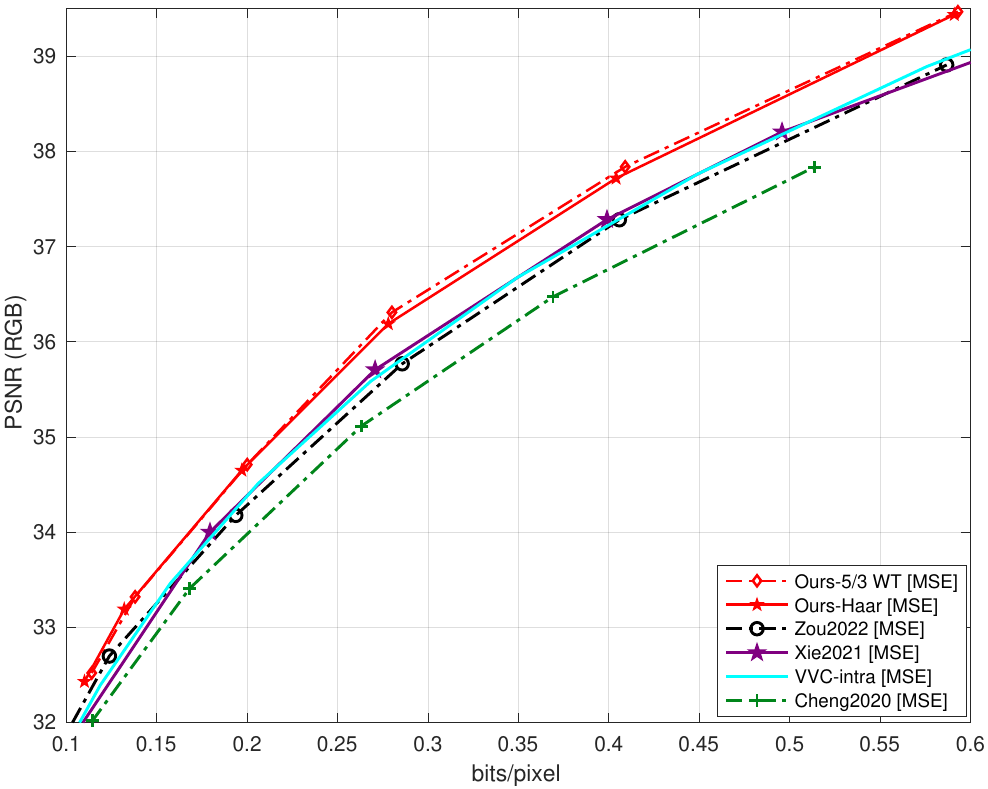}
    \caption{}
    %\label{fig:clic-b}
  \end{subfigure}
  \caption{(a) The average PSNR performances of different methods in the Tecnick dataset. (b) The average PSNR performances of different methods in the CLIC dataset.}
  \label{fig:tecnick-clic}
\end{figure*}

Fig. \ref{fig:tecnick-clic} (a) reports the PSNR performances of different methods in the Tecnick 100 dataset. Among the PSNR-optimized methods, Xie2021 \cite{xie2021enhanced}  achieves the best performance in other compared methods. Our method with Haar transform also outperforms Xie2021 \cite{xie2021enhanced} by about $0.2$ dB at most rates, and achieves $-9.46\%$ BD-Rate reduction over VVC. Our method with 5/3 wavelet can further improve up to 0.1 dB.

Fig. \ref{fig:tecnick-clic} (b) compares the PSNR performances of different methods in the CLIC 2021 test set. Our method with Haar transform has even more gains over Xie2021 \cite{xie2021enhanced}, Zou2022 \cite{Zou_2022_CVPR}, and VVC, with up to $0.5$ dB at high rates. Its BD-Rate reduction over VVC is $-9.20\%$.  Our method with 5/3 wavelet is also better than the Haar transform slightly.

\subsection{Performance and Speed Trade-off}

\begin{table*}[!t]
\caption{Comparisons of encoding/decoding time, BD-Rate reduction over VVC, and model sizes of the low bit rates and high bit rates for Kodak test set.}
\begin{center}
\begin{tabular}{ccccc}
\hline
\textbf{Methods} & \textbf{Enc. Time} & \textbf{Dec Time} & \textbf{BD-Rate} &\textbf{$\#$Params}\\ 
\hline
  VVC  & 402.3s& 0.61s& 0.0 & -\\
  
  %He2022 \cite{He_2022_CVPR} &0.254s &0.215s &-6.24 \% & 10.10MB\\
  Cheng2020 \cite{cheng2020}  &27.6s& 28.8s& 2.6 \% & 50.80 MB \\
  Hu2021 \cite{Hu_2021}  &32.7s& 77.8s& 11.1 \% & 84.60 MB \\
  He2021 \cite{He_2021_CVPR}  &20.4s& 5.2s& 8.9 \% & 46.60 MB \\ 
  Xie2021 \cite{xie2021enhanced}&4.097s &9.250s&-0.8 \% & 128.86 MB\\
  Zhu2022 \cite{zhu2022transformerbased} &0.269s &0.183s &-3.9 \% & 32.34 MB\\
  Zou2022 \cite{Zou_2022_CVPR} &0.163s &0.184s &-2.2\% & 99.86 MB\\
  Qian2022 \cite{Qian2022_ICLR}&4.78s &85.82s&3.2 \% & 128.86 MB\\
  Fu2023 \cite{GLLMM}  &420.6s& 423.8s& -3.1\% & - \\ 
  \hline

\textbf{Baseline} & \textbf{0.109s}  & \textbf{0.142s} & \textbf{-0.4\%} & \textbf{52.22 MB} \\
\textbf{Baseline+WeConv(Haar)}  &\textbf{0.110s}& \textbf{0.147s}& \textbf{-4.7\%} &  \textbf{58.41 MB}\\
\textbf{WeConvene(Haar)}  &\textbf{0.352s}& \textbf{0.388s}& \textbf{-8.2\%} &  \textbf{107.15 MB}\\ 
\textbf{WeConvene(5/3 WT)}  &\textbf{0.363s}& \textbf{0.415s}& \textbf{-9.4\%} &  \textbf{109.23 MB}\\ 
\textbf{WeConvene(9/7 WT)}  &\textbf{0.386s}& \textbf{0.445s}& \textbf{-9.8\%} &  \textbf{113.46 MB}\\ 
   \hline%%%FIRST BIT RATE
\end{tabular}

\label{runing_time}
\end{center}
\end{table*}

Table \ref{runing_time} compares the average encoding/decoding times, BD-rate reductions over VVC \cite{BDRate}, and the number of model parameters (obtained by the PyTorch Flops Profiler tool) of various methods on the Kodak test set, using a NVIDIA Tesla V100 GPU with 12 GB memory, except for VVC, which only runs on CPU (a 2.9GHz Intel Xeon Gold 6226R CPU is used). The number of parameters of \cite{GLLMM} is not available, since it is written in TensorFlow, but it is shown in \cite{GLLMM} that its model complexity is much higher than \cite{cheng2020}.

To study the contributions of WeConv and WeChARM separately in our method, we design a simplified baseline scheme for our method by removing the DWT/IDWT in WeConv and WeChARM, and only using one ChARM module in entropy coding. On top of the Baseline, we enable the WeConv and then the two-step WeChARMs. In each case, we retrain the entire system to get its best performance. Table \ref{runing_time} includes results of our Baseline, Baseline + WeConv (Haar), and the full WeConvene with Haar, 5/3, and 9/7 wavelets respectively.

The decoding time and BD-Rate reductions of some methods are also reported in Fig. \ref{BD_rate} earlier.

The encoding and decoding times of learned methods \cite{cheng2020, GLLMM, xie2021enhanced, Qian2022_ICLR} are relatively slow because they employ sequential entropy models and cannot be accelerated by GPU. Some recent LIC approaches such as \cite{zhu2022transformerbased, Zou_2022_CVPR} are much faster, by using GPU-friendly parallelizable entropy models. Their R-D performances are also among the best.

The BD-Rate reduction of the proposed WeConvene scheme with 9/7 wavelet is $5.9\%$ better than \cite{zhu2022transformerbased}, and $-9.8\%$  better than VVC, making our method the new state of the art. Our model complexity with different wavelets is only $8-14\%$ higher than \cite{Zou_2022_CVPR}. The encoding/decoding time is about twice of \cite{Zou_2022_CVPR}. This is mainly because we use two sequential WeChARM modules in the entropy coding part, but our method is still significantly faster than many other LIC methods.

\subsection{Contributions of Different Modules in WeConvene}
\label{Ablation}

\begin{figure*}[tb]
  %\centering
  \begin{subfigure}{0.48\linewidth}
    \includegraphics[scale=0.335]{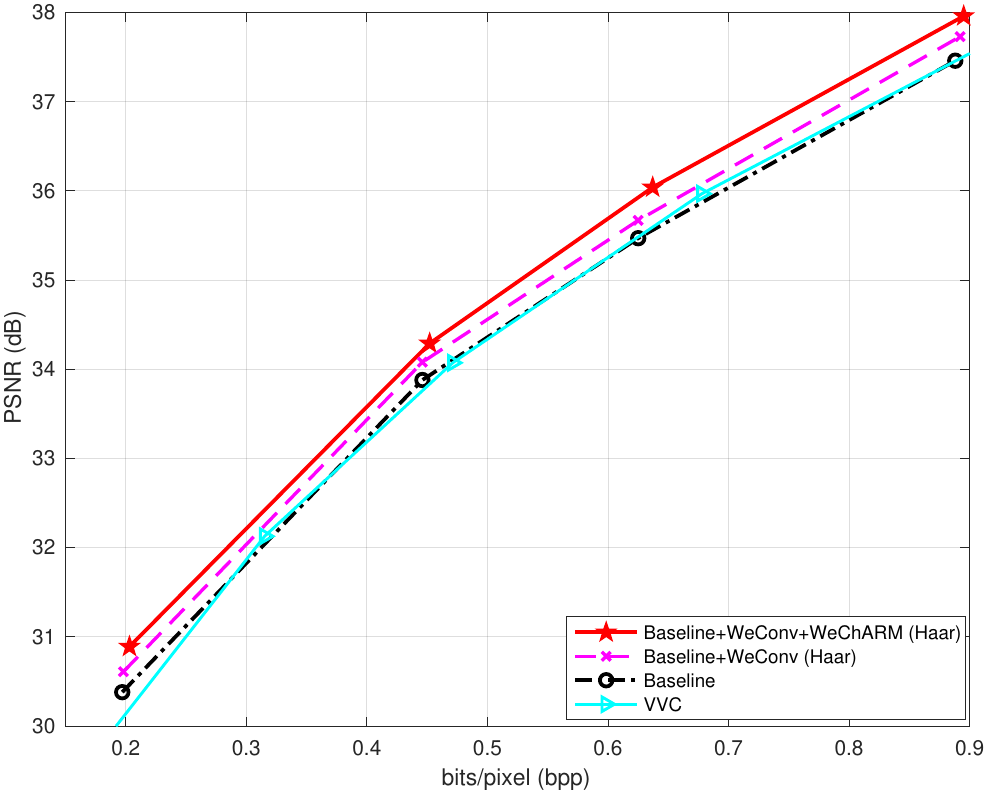}
    \caption{}
    %\label{abalation}
  \end{subfigure}
  \hfill
  \begin{subfigure}{0.50\linewidth}
    \includegraphics[scale=0.34]{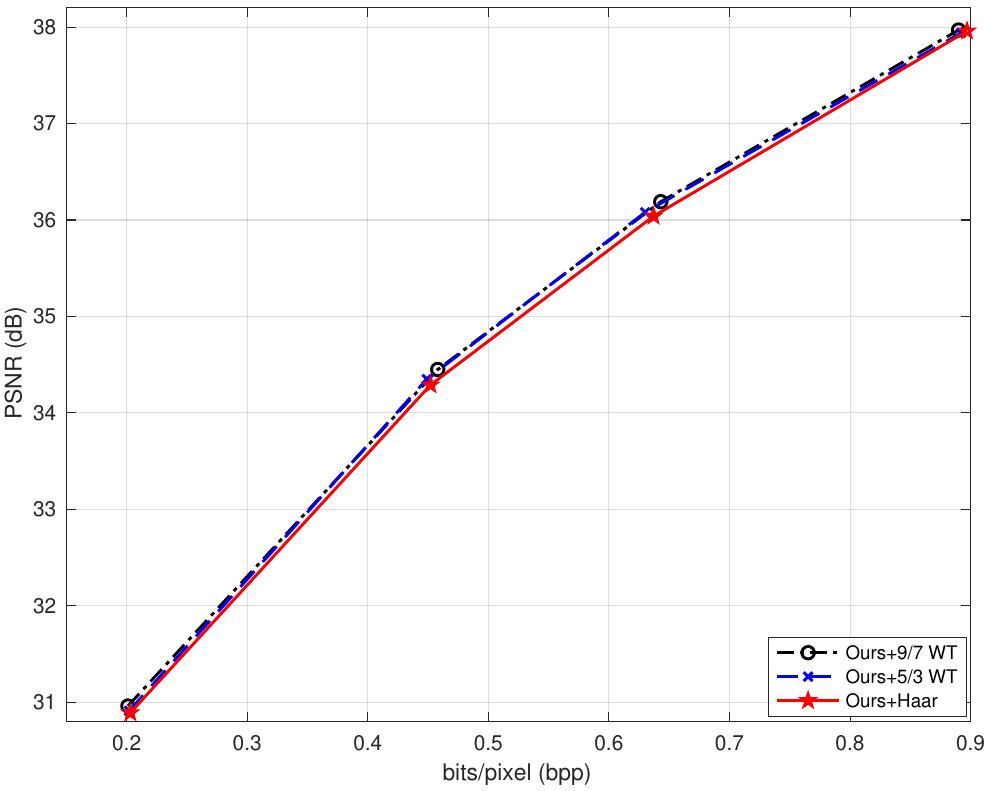}
    \caption{}
    %\label{Different_wavelets}
  \end{subfigure}
  \caption{(a) R-D performances of VVC and different configurations of our method for the Kodak dataset using the Haar transform. (b) R-D performances of VVC and different wavelets (Haar, 5/3, and 9/7 wavelets) in WeConvene for the Kodak dataset.}
  \label{abalation}
\end{figure*}

% \begin{figure*}[!thp]
% 	\centering
% 		\includegraphics[scale=0.5]{./figures/Kodak_abalation.pdf}
% 	\caption{R-D performances of VVC and different configurations of our method on the Kodak dataset.}
% 	\label{abalation}
% \end{figure*}

Table \ref{runing_time} includes the performances of our Baseline and Baseline + WeConv (Haar). Both of them are faster than \cite{zhu2022transformerbased, Zou_2022_CVPR}. The BD-rate reduction of the Baseline over VVC is only $-0.4\%$. Enabling WeConv with the Haar transform almost does not increase the encoding/decoding time, but it can achieve an impressive $-4.7\%$ BD-rate reduction over VVC, which is already better than other LIC methods in the table. The model complexity is only increased by about $10\%$ compared to the Baseline.

Fig. \ref{abalation} (a) compares the R-D curves of VVC and different configurations of our scheme on the Kodak dataset using the Haar transform. It can be seen that our baseline achieves similar performance to VVC. When WeConv is enabled, the performance is improve by about 0.2 dB at all rates. When WeChARM is also enabled, another gain of 0.2 dB can be achieved.

\subsection{Contributions of Different Wavelets}
\label{differnet_Wavelets}

% \begin{figure*}[!thp]
% 	\centering
% 		\includegraphics[scale=0.4]{./figures/different_wavelets.pdf}
% 	\caption{R-D performances of VVC and different Wavelets on the Kodak dataset.}
% 	\label{Different_wavelets}
% \end{figure*}

In this experiment, we replace the Haar wavelet with the 9/7 and 5/3 wavelet. Other configurations remain the same. The experimental results are shown in  Fig. \ref{abalation} (b). The 5/3 wavelet improves performance by about 0.05-0.1 dB at the same bit rate compared to the Haar. The 9/7 wavelet has almost the same performance as the 5/3 wavelet. The reason is that the input sizes to the WeConv modules are not very large in this paper.

\subsection{Comparison of Different Channel Slices in WeChARM}
\label{Ablation}

\begin{table}[!thp]
\caption{The performance of different channel slices}
\begin{center}
  \begin{tabular}{ccccccc}
  \hline
 \textbf{Groups} & \textbf{Bit rate}  & \textbf{PSNR } & \textbf{MS-SSIM }& \textbf{Enc. time} & \textbf{Dec. time}& \textbf{$\#$Params}\\
  \hline
  \textbf{5} & 0.162  & 30.12 dB & 12.78 dB  &0.352 ms &0.388 ms &107.15  MB \\
  \textbf{10} & 0.167  & 30.16 dB & 12.82 dB &0.424 ms &0.491 ms &179.09 MB\\
  \hline
  \textbf{5} & 0.894   & 37.96 dB & 20.53 dB &0.352 ms &0.388 ms &107.15 MB\\
  \textbf{10}& 0.9023  & 38.01 dB &20.57 dB  &0.424 ms &0.491 ms &179.09 MB\\
  \hline
\end{tabular}
\label{Tab:abalation_spilt}
\end{center}
\end{table}

Table \ref{Tab:abalation_spilt} studies the impact of the number of channel slices in the channel-wise entropy coding when the Haar transform is used. Results with 5 and 10 slices at low rate and high rate are reported.

It can be observed that at both low rate and high rate, when the latent representations are divided into 10 slices instead of 5 slices, the R-D performance only increases slightly. On the other hand, the model size increases about 67\%, and the encoding/decoding time increases 20-25\%. This is because when there are too many slices, the number of channels is smaller in each slice, making it less efficient to reduce the redundancy. Moreover, since the slices need to be coded sequentially, the encoding/decoding time is also increased. Therefore we choose to use 5 slices in WeChARM.

\section{Conclusions}

This paper introduces a simple but efficient approach to use wavelet transform in both the convolution layers and entropy coding of the learned image compression (LIC). It makes the latent representations sparser in wavelet domain, which helps to achieve better R-D performance. 

For the Kodak dataset and the baseline network with $-0.4\%$ BD-Rate saving over H.266/VVC, introducing WeConv with the simplest Haar transform improves the saving to $-4.7\%$. This is quite impressive given the simplicity of the Haar transform. Enabling Haar-based WeChARM entropy coding further boosts the saving to $-8.2\%$. When the Haar transform is replaced by the 5/3 or 9/7 wavelet, the overall saving becomes $-9.4\%$ and $-9.8\%$ respectively. The complexity of the scheme is also significantly lower than most LIC methods.

The framework in this paper opens up many future research topics, and allows the rich theories and results in the wavelet community to be introduced to learned image/video coding. For example, multiple levels of wavelet transforms can also be employed. Another possible approach is to use different wavelets in different layers, e.g., longer wavelets when the input size is larger, and shorter wavelets when the input is smaller.

In addition, as a standalone convolution layer module, the WeConv can also be used in many other computer vision tasks beyond image/video compression.

%\clearpage  % TODO REVIEW/FINAL: This \clearpage needs to be removed from both review and camera-ready versions.

\bibliographystyle{splncs04}
\bibliography{main}
\end{document}